# Anisotropic wetting and de-wetting of drops on substrates patterned with polygonal posts

**Robert J. Vrancken[†], Matthew L. Blow[‡,\*], Halim Kusumaatmaja[§], Ko Hermans[†], An M. Prenen[†], Cees W. M. Bastiaansen[†,°], Dirk J. Broer[†] and Julia M. Yeomans[‡]**

[†]*Laboratory of Functional Devices, Department of Chemical Engineering and Chemistry, Eindhoven University of Technology, P.O. Box 513, 5600 MB Eindhoven, The Netherlands,*
[‡]*The Rudolf Peierls Centre for Theoretical Physics, Oxford University, 1 Keble Road, Oxford OX1 3NP, U.K.,*
[\*]*Centro de Física Teórica e Computacional, University of Lisbon, Avenida Professor Gama Pinto 2, Lisbon P-1649-003, Portugal,*
[§]*University Chemical Laboratories, University of Cambridge, Lensfield Road, Cambridge CB2 1EW, U.K.,*
[°]*School of Engineering and Materials Science, Queen Mary University of London, Mile End Road, London E1 4NS, UK*

**Table of content**

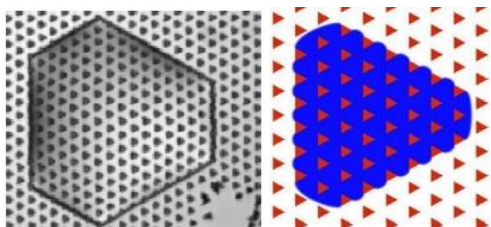

We record contact line motion of ink-jet printed drops spreading and evaporating on surfaces patterned with polygonal microposts and, aided by lattice Boltzmann simulations, explain the drop shapes in terms of interface pinning at posts.

## Abstract

We present results showing how water drops, produced by ink-jet printing, spread on surfaces patterned with lattices of diamond or triangular posts. Considering post widths typically ~7 μm and lattice spacings between 15-40 μm, we observe drop shapes with 3,4 and 6-fold symmetry, depending on both the symmetry of the lattice and the shape of the posts. This is a result of the different mechanisms of interface pinning and depinning which depend on the direction of the contact line motion with respect to the post shape. Lattice Boltzmann simulations are used to describe these mechanisms in detail for triangular posts. We also follow the motion of the contact line as the drops evaporate showing that they tend to return to their original shape. To explain this we show that the easy direction for movement is the same for spreading and drying drops. We compare the behaviour of small drops with that of larger drops created by jetting several drops at the same position. We find that the contact line motion is unexpectedly insensitive to drop volume, even when a spherical cap of fluid forms above the posts. The findings are relevant to micro-fluidic applications and to the control of drop shapes in ink-jet printing.

## 1. Introduction

Driven by developments in surface patterning methods, there have been significant academic interests in recent years to control the spreading and flow of liquids using the surface properties of the solids with which they are in contact. It is now possible to produce complex surfaces with regions or patches of varying compliances, wettabilities and/or topological features.[1–10] These structured surfaces provide useful model systems to explore fundamental issues such as contact angle hysteresis[11–13] and three-phase contact line motion.[14–18] Understanding how liquids move on structured surfaces also helps to develop many practical applications. In particular, one rapidly growing area of applications is micro-fluidics. For example, chemically patterned surfaces may be exploited to generate monodisperse drops[19] and numerous superhydrophobic surfaces have been fabricated to reduce drag and contact angle hysteresis.[20,21]

Topographical patterning may also be utilised to make droplet spreading or motion across a surface anisotropic,[22] for example using ratcheted posts[23–25] or bent hairs.[26,23] In this paper, our focus is to investigate the movement of the contact line of a fluid on surfaces patterned with polygonal posts. Courbin et al.[27,28] showed that the final shape of a drop spreading on a surface patterned with posts can strongly depend on the arrangement of the posts: square drops are obtained when the posts are in a square lattice, and hexagonal drops when the posts are in a triangular lattice. The shape of the posts was not explored in the study of Courbin *et al.*, but it was later investigated in experiments[29] and lattice Boltzmann simulations.[30,31] It was found that the post shapes indeed influence the drop shape due to the anisotropy in the way in which the fluid de-pins from the posts locally. Two distinct de-pinning mechanisms were described for a fluid imbibing this type of surface. Here we experimentally validate the existence of anisotropic wetting on surfaces with lattices of polygonal posts. Furthermore, we extend the analysis to the retracting motion of the contact line (i.e. employing 'reverse imbibition' or drying), which also exhibits two de-pinning mechanisms. Interestingly we find that the preferential directions for spreading and retraction are the same for the surfaces investigated.

For our purposes a suitable experimental method for depositing the drops is inkjet printing, which is increasingly used as a versatile research tool.[32,33] Here it allows us to deposit one or multiple drops on the patterned surfaces in a very controlled fashion. In this way we are able to quickly create a large number of drops with near identical conditions, in order to study small drops across multiple surfaces as well as obtain high reproducibility on a single surface, even for pico-liter droplets. A further advantage is being able to vary the total drop volume by printing multiple drops on one location of the surface, thereby looking at the evolution of the drop shape (in particular the contact line) with increasing volume.

The experimental results are compared to lattice Boltzmann simulations. The lattice Boltzmann method is a mesoscale simulation technique which has proved to be successful in predicting and understanding numerous wetting situations.[34–38] The particular model we use is detailed in [39,40] (see also the supporting information). Alternative models are also available in the literature, see e.g. [41–45].

This paper is structured as follows: after a description of the experimental details of inkjet printing and surface preparation in section 2, an analysis of the various shapes of the advancing contact line for drops inkjet-printed onto variously patterned surfaces is presented in section 3. Also included in this section is a comparison to our previous lattice Boltzmann simulation results, and a detailed description of the de-pinning mechanisms on a surface patterned with triangular posts arranged on a hexagonal lattice. In section 4 the analysis is extended to retracting contact lines via evaporation experiments and lattice Boltzmann simulations, together with a theoretical analysis of the de-pinning mechanisms of the contact line. Finally in section 5 the main conclusions are briefly presented.

## 2. Experimental methods

Printing was performed using a MicroDrop inkjet printing system with a MD-K-140 glass capillary and nozzle of 98 μm diameter (MicroDrop GmbH, DE). The electrical pulse supplied to the piezo-element was 100 V, with a 140 μs pulse length. Deionised water was used as printing fluid, which created highly reproducible drops of around 200 pl, ejected from the nozzle at 6 ms$^{-1}$, as was determined by the integrated camera system and software of the labscale inkjet printer. Substrates (see below) were placed under the inkjet nozzle in a fixed position, and a pre-set amount of drops, up to 225, was then printed onto the surface, forming a single larger drop on the surface. Print frequencies were varied from 1 to 200 Hz, so within a little over 1 second even the largest surface drops were completely formed. The stand-off distance between the nozzle and the surface was 1 mm.

The surfaces were created as in our previous work by means of photo-lithography of negative photoresist SU-8 (MicroChem Corp.).[35] First, a thin layer of SU-8 (3 to 5 μm) was spin-coated from a solution of cyclopentanone onto standard laboratory glass slides, which were previously cleaned in ethanol and subsequently dried in nitrogen. Secondly a heating sequence on a hotplate was performed (1 min at 65 °C followed by 2 min at 95 °C) to remove the solvent, while the sample was simultaneously flood exposed with UV light (Exfo UV source) to promote adhesion of the SU-8 surface texture. Subsequently a second layer of SU-8 with a thickness of 18 μm was again applied and dried without exposure to flood-UV. Then, the SU-8 was exposed to UV-light (Exfo UV source) via contact lithography with a quartz-chrome custom lithographic mask. The mask was patterned with various post geometries (round, square, triangle, hexagon) and two lattice types (hexagonal, square) at various post sizes and lattice spacings, while the post height remained fixed by the layer thickness of 18 μm. Following exposure, a second heating sequence (1 min at 65 °C followed by 2 min at 95 °C) was employed to crosslink the exposed areas of the layer, followed by slow cooling. The non-exposed parts were then removed by a custom developer (mr-dev 600, MicroChem Corp.), after which the sample was rinsed by isopropanol and blown dry with a mild nitrogen flow. Reference samples of SU-8 with a smooth flat surface were created via the same procedure without the second patterning step. The intrinsic contact angle of this smooth surface of SU-8 was measured by contact angle measurement setup (OCA-30, DataPhysics Germany) to be 63±3°. The advancing (65±3°) and receding (25±5°) contact angles were also determined via automatically adding and retracting liquid with the syringe-needle of the setup inserted. No noticeable swelling was observed during either the contact angle measurements or during the microscopy experiments, where this was monitored closely.

After printing, the drops were imaged by optical microscopy using a Leica DM6000 M (20x objective with 1x additional magnification), equipped with a Leica DFC420C camera in bright field transmission mode. Images from the movie were exported as JPEG files via VirtualDub video editing freeware (version 1.9.0) and subsequently analysed.

The sample with water drop printed on top was moved to the microscope manually with care, within 20 to 30 seconds after printing is completed. The drop and surface were subjected during movement to minor mechanical vibrations. The impact of these mechanical vibrations was found to be negligible, as reproducibility of the drop shape during several identical experiments was found to be within experimental error. Similarly, by moving the same sample with drop multiple times, no differences in drop shape were observed. One reason for this is the fact that after print-ing, the drops were in the collapsed ('Wenzel') state,[46] which meant that a large portion of the total drop interface was in contact with the surface and such Wenzel states are known to be quite immobile due to contact line pinning and hysteresis.

After being placed under the microscope, the drops started to evaporate considerably faster than under ambient conditions, due to the heat of the additional transmitted light. While this heat flux was not strictly regulated, it was reproducible between measurements as long as the optics of the microscope (diaphragm, light intensity, magnification) were not changed. Typical evaporation times were between 30 s and 5 min, depending on drop volume. This increased evaporation was beneficially utilised in the experiments with receding contact lines.

## 3. Advancing drops and films

### 3.1. Drop shape anisotropy

Drops which are inkjet-printed on surfaces patterned with polygonal posts exhibit a variety of typical shapes, depending on the lattice type, post shape and size. The geometry is described by post height $h$, post width $b$ and lattice spacing $d$. In this section we give a qualitative description of the factors influencing the drop shapes, before presenting a detailed description of the mechanisms that determine the shapes for triangular posts.

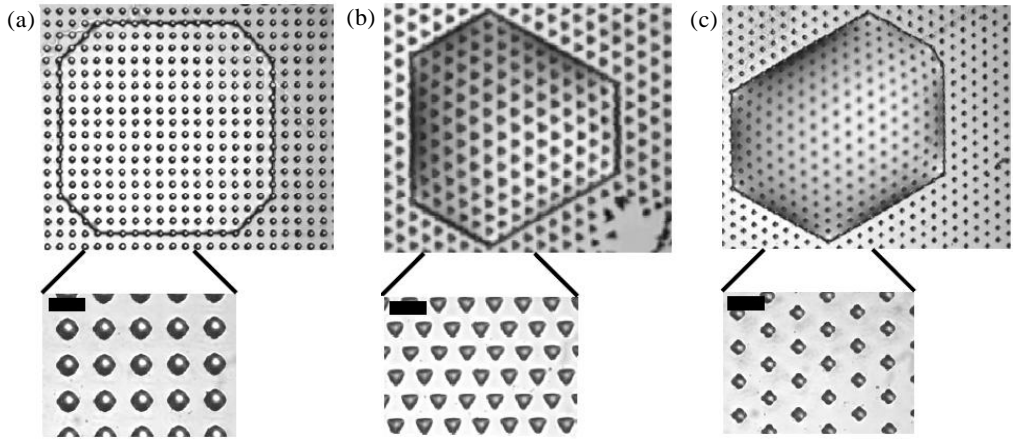

**Fig. 1** Microscopic images of various drop shapes on different types of lattice and post shapes, with a magnification of the posts for each surface geometry. The black scale bar indicates $20\ \mu m$. (a) An octagonal drop on a square lattice with diamond shaped posts ($b = 10\mu m, d = 20\mu m$). (b) A drop with a shape intermediate between a hexagon and an equilateral triangle on a hexagonal lattice with triangular posts ($b = 7.5\mu m, d = 15\mu m$). (c) An irregular hexagon on a hexagonal lattice with diamond shaped posts ($b = 7.5\mu m, d = 20\mu m$). Due to curvature of the liquid-vapour interface, the posts below the drop in (b) and (c) look slightly distorted and out of focus, whereas the drop in (a) was imaged after prolonged evaporation just prior to initial retraction of the contact line (due to depinning) and therefore had a nearly horizontal liquid-vapour interface.

In figure 1, a number of typical shapes are shown. From studying the various shapes of the drops in figure 1, and other drops on a variety of surface geometries, it is apparent that the anisotropy in the spreading can result from both the lattice in which the posts are arranged, and the shape of the posts themselves. The effect of lattice type is clearly illustrated in figure 1(a), with the contact line closely conforming to the square lattice of posts, but with the corners slightly clipped to produce an octagon shape. The anisotropy effect of the lattice can be explained by the contact line being pinned to rows of posts, and having to overcome an energetic penalty in order to advance from one row to the next.[27,30] When it is finally able to reach such a new row, the adjacent posts are quickly wet, akin to the unzipping mechanism identified by Sbragaglia *et al.*[36], as well as our previous work on surfaces with parallel corrugations, where the contact line occasionally 'jumps' to the next ridge and then proceeds to wet this new corrugation along its length.[37]

Likewise, figures 1(b) and (c) show the influence of a hexagonal lattice, but they are also clearly affected by the shape of the posts. Comparing 1(b) and 1(c) is instructive. In 1(b) the post symmetry is reinforced by the lattice symmetry, of which it is a rotational subgroup. The post symmetry influences the shape of the interface considerably, as the contact line is able to reach the next row of posts more easily in the direction of the sharp edges of the triangular pillars. In 1(c), however, the post and lattice symmetries 'clash', with one not being a subgroup of the other. This leads to a less distinct difference in wetting directions, as no particular direction is clearly favourable over other directions. We expect the resulting shape to have 2-fold symmetry, this being the common factor of the post and lattices symmetries, but in practice this symmetry is only approximate, as illustrated in 1(c). We further note that circular posts arranged on a hexagonal lattice were used in the experiments by Courbin *et al,.*[27] and in that case regular hexagons (which possess the lattice 6-fold rotational symmetry) were observed.

The symmetry of the drop in figure 1(c) is not perfect and this is usually the case in the experiments. It can be at-tributed to either mechanical vibrations, local deviations from ideal shapes of the posts and other minor influences such as drops ejected by the inkjet nozzle under an angle ('satellite drops' or 'side shooters'). A small perturbation that causes the interface to reach the next row of posts first in a particular direction can have a visible effect on the drop shape as the unzipping along that row of posts will follow very quickly. However, the reproducibility of the drop shapes is generally quite high, and it can, as we demonstrate later, be analysed statistically.

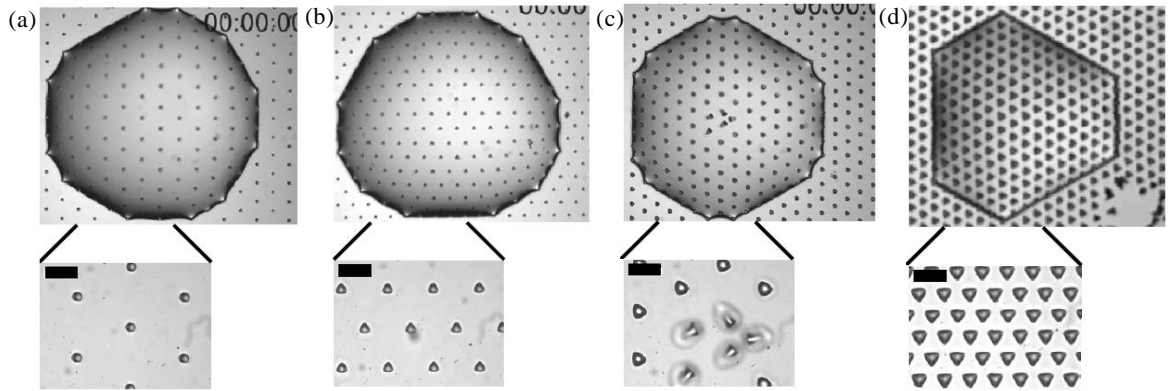

**Fig. 2** Microscopic images of various drop shapes on a hexagonal lattice with differently shaped and spaced posts, with a magnification of the posts for each surface geometry. The black scale bar indicates $20\ \mu m$. (a) A nearly round drop on a sparse lattice of small round pillars ($b = 5\mu m, d = 40\mu m$). (b) A drop with a shape intermediate between sphere, a hexagon and a triangle (pointing upward), on a sparse lattice of small triangular posts ($b = 7.5\mu m, d = 30\mu m$). (c) A drop with an approximately hexagonal shape on an intermediately filled lattice with triangular posts ($b = 10\mu m, d = 30\mu m$). (d) A drop shaped intermediately between a regular hexagon and a triangle, on a dense lattice with relatively thick triangular posts ($b = 7.5\mu m, d = 15\mu m$). Note that figure 2(d) is identical to figure 1(b).

To further display the influence of the shape and arrangement of the posts, several different examples of posts arranged on a hexagonal lattice are presented in figure 2. The final shapes have a complicated dependence on both the lattice type and spacing and on the geometric parameters of the substrate $h$, $b$ and $d$. However, in general when the posts are small compared to the lattice spacing and relatively far apart as in figure 2(a), the contact line more closely resembles a circle. This is because the surface tension of the drop interface dominates over the pinning effects of the posts.

**3.2. Simulation results**

Previous lattice Boltzmann simulation results and analysis for advancing imbibition were presented in [30,31]. We will first briefly revisit these findings for the purpose of comparison with the deposition experiments presented in this paper. The details of the simulation method are given in the online Supporting Information.

Simulating a large drop feeding imbibition among an array of posts beneath it is very computationally expensive. Since we are only interested in the details of flow amongst the posts, it is expedient to replenish the drop from a 'virtual reservoir', a small circular region of radius ∼6 lattice units at the centre of the box where additional mass of liquid is introduced. Using this simulation geometry, we were able to capture a number of anisotropic spreading behaviours. For example, of particular relevance here, for a hexagonal array of triangular posts, the spreading liquid quickly facets into a hexagon, which then forms a drop shaped intermediate between a regular hexagon and a triangle as shown in figure 3(a)-(c). Note that the final drop shape in figure 3(c) is very similar to our experimental observation in figure 1(c).

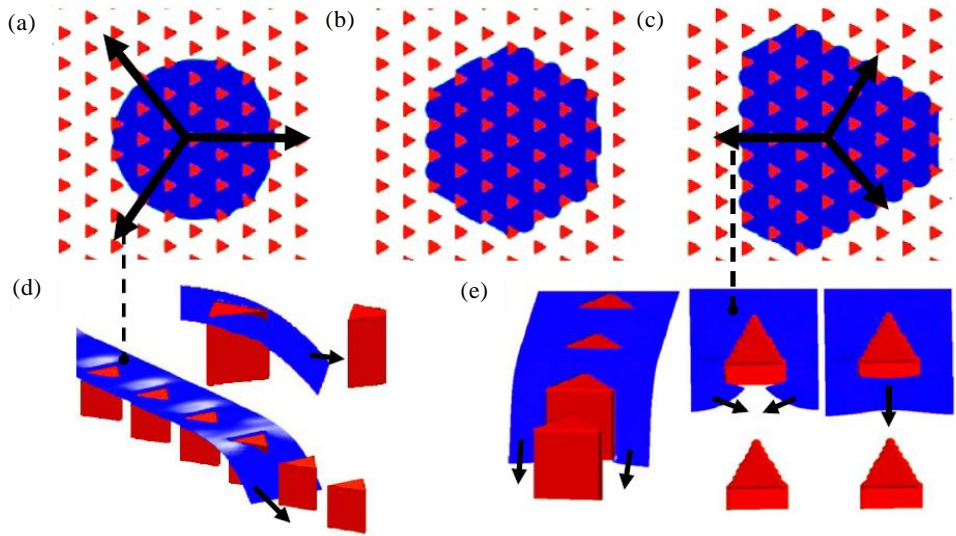

**Fig. 3** Lattice Boltzmann simulations of spreading drops on a surface covered with triangular posts (*with b* = 20, *d* = 40 *and h* = 30) in a hexagonal array. The intrinsic contact angle is 55°. These results were reported previously in [31]. (a) – (c) Drop shape as function of time, showing the evolution of the drop from approximately spherical to hexagonal to a shape intermediate between hexagonal and triangular. Side view of the (d) connected contact line (CCL) and (e) disconnected contact line (DCL) de-pinning mechanisms. The movement of the contact line is indicated with arrows. The simulations are periodic in a direction perpendicular to the motion of the interface. The dashed lines link the two de-pinning mechanisms with the corresponding lattice directions indicated in (a) and (c) with thick black arrows.

To understand the spreading anisotropies, we analysed the ways in which the contact line de-pins. We found two distinct routes which we shall call the connected and disconnected contact line mechanisms, indicated in figure 3(d) and (e) respectively. The simulations reported in these figures are periodic in a direction perpendicular to the motion of the interface. Consider first figure 3(d) where the interface reaches a flat face of the triangular post first. The film is of height $h$ at the back face of the leading triangle, completely wetting this face. Beyond this, the height of the film decreases, such that the interface meets the substrate with the Young angle $\delta$. The interface will remain pinned if this condition can be geometrically satisfied without the contact line reaching the next row of posts. Hence, if the interface is flat, pinning will occur if $\delta > \tan^{-1} h/d$,[27] although we showed in [31] that the curvature of the interface needs to be taken into account to obtain an accurate prediction. Because the contact line is unbroken, this method of pinning will be called the connected contact line mechanism ('CCL').

On the other hand, if the liquid reaches a sharp edge of the triangular posts first, pinning can occur with the interface spanning the gaps between the post faces, as show in figure 3(e). In this situation, the contact line along the base substrate is punctuated by these faces, so we use the term disconnected contact line mechanism ('DCL'). Whether there will be pinning or contact line advance depends on the balance of the interfacial free energy cost of creating interface against the free energy reduction of wetting the hydrophilic substrate. The latter will outweigh the former when $\delta$ is sufficiently small. As $\delta$ is lowered, or because of the inertia of the fluid motion, the interface creeps around the corners of the post face. Once it comes into contact with the neighbouring portion of interface, it readily wets up the post face and across the base substrate towards the next row of posts.

We note that Jokinen *et al.*[29] propose that the connected contact line mechanism occurs in both directions. However, it is worth noting that they study posts which are elongated triangles rather than equilateral, and which are more closely spaced than considered by us.

The experimental variations in drop shape in figures 1 and 2 can be explained well in terms of the analysis of the lattice, combined with the CCL and DCL de-pinning mechanisms. In general the threshold values of $\delta$ or fluid inertia, for the two pinning mechanisms will differ, giving rise to an intermediate range of $\delta$ in which spreading is permitted only in certain directions, and is thus highly anisotropic. For example, in Fig 2(d) the contact line depins more easily in directions where it reaches the flat faces of the triangular posts first i.e. the CCL route dominates. Qualitatively similar shapes, intermediate between an equilateral triangle and a regular hexagon, can be observed by comparing the simulation result of figure 3 with the experimental results shown in figure 2(d). We note, however, that there are significant differences in the circumstances represented: 2(d) shows a Wenzel drop having a dome-shaped cap extending above the posts, which has impacted onto the substrate with substantial kinetic energy, while 3(c) shows an imbibed film (a thick, flat film with the same height as the posts, but not wetting their top faces) which has been introduced quasi-statically. We therefore now describe further experiments to investigate how the final drop shape depends on details of how it is produced.

### 3.3. Reproducibility of drop shapes from inkjet printing

We now investigate how the final shape of the drop changes with its volume and, in particular, assess the reproduci-bility of the drop shapes. The different volumes were created by the rapid sequential addition of between 1 and 225 drops of volume 200 pl each (see section 2). Our results are for the surface with triangular posts on a hexagonal lattice presented in figure 1(c). The experimental surface parameters are $h = 18$ μm, $b = 10$ μm (the sides of the triangular posts have length 10 μm) and $d = 15$ μm.

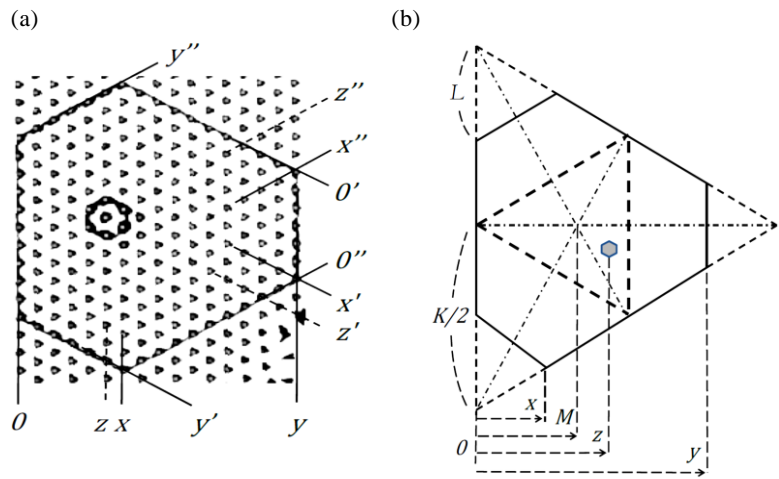

**Fig.4** Parameters used in the shape analysis of the hexagonal drops. (a) Two superimposed screenshots of the contact line of a drop, the outer initial outline of the drop, and the inner last position before it is completely evaporated. The positions $x$, $y$ and $z$ are natural numbers of rows of the lattice, as defined in the figure and counted from the left initial boundary. The position $z$ is the middle post of the last remaining part of the drop. Because of the approximate threefold rotational symmetry of the shape, positions $x'$ and $x''$ can be defined similarly to $x$, and equally for $y$ and $z$. (b) An equilateral triangle with side $K$, showing observables $x, y$ and $z$ and the position of the centroid $M$. Also shown is the length $L$ by which the side of the hexagon would have to be extended to form an equilateral triangle.

The drops form truncated triangles. To characterise these we measure the lengths $x, y, z$ which are defined in figure 4 as the number of rows of posts between the left-hand edge of the drop $O$ and the lowest corner $x$, right-hand edge of the drop $y$, the location of the final wetted part of the drop during evaporation $z$ and the centroid of the drop $M$, respectively (see figure 4(b)). In general, because the truncated hexagons formed in experiments are not perfectly regular, the measured values of $x, y, z$ will be different if the origin is chosen with the drop rotated by $120^0$ or $240^0$. In the figure, and data presented below, this is indicated by also using parameter sets $x', y', z'$ and $x'', y'', z''$ for a given drop. Measurement of the three sets of lengths for each drop allows us to fully utilise the data and to check for systematic errors.

It is useful to define the dimensionless parameter

$$\lambda = \frac{2x}{x+y}, \tag{1}$$

and, similarly, $\lambda'$, $\lambda''$. $\lambda$ is a measure of the degree to which the triangle circumscribing the hexagon is truncated. The meaning of this parameter is illustrated in figure 4(b) which shows a regular truncated triangle. $K$ is the length of the side of a triangle and $L$ the length of the equilateral triangle cut off the corners. For the regular shape $\lambda = 2L/K$. Hence $L=0$, corresponding to $\lambda=0$, denotes an equilateral triangle. Furthermore $L=K/3$ or $\lambda=2/3$ indicates a regular hexagon, and $L=K/2$ or $\lambda=1$ a triangle pointing in the opposing direction to the posts.

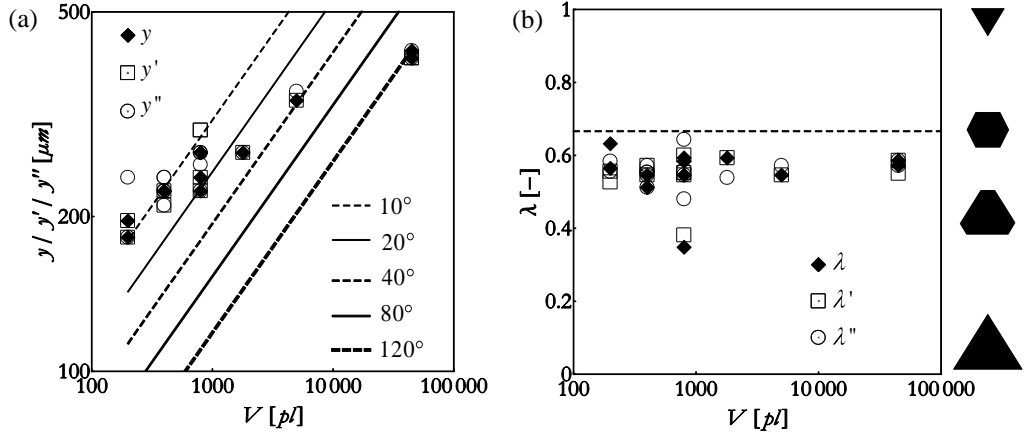

**Fig. 5** (a) Drop sizes $y, y', y''$ measured in the three principal lattice directions as a function of drop volume $V$ plotted on double logarithmic scale. Also plotted is the theoretical curve of the drop diameter for a spherical drop with an intrinsic contact angle $\delta$ as defined in the legend, calculated using equation (2). (b) Drop shape parameters $\lambda, \lambda', \lambda''$ for the three principal lattice directions as a function of $V$, plotted on log-linear scale. Note that the three lowest measured points with $\lambda < 0.5$ are from one single observed drop which more closely resembled a triangle than all other drops. The drop shapes corresponding to their equivalent $\lambda$ are indicated to the right of (b), where $\lambda = 2/3$ indicates a regular hexagon (plotted as a dashed line).

Figure 5 demonstrates the effects of drop volume on drop shape and size. The data for $\lambda$ and $y$ are equally distributed for all three principal lattice directions, reflecting the 3-fold symmetry of the drops. Figure 5(a) shows that the linear size of the contact line varies little as the volume of the drops is increased, indicating strong contact line pinning and hence large contact angle hysteresis. The theoretical curves for the radius of a drop forming a spherical cap for a given volume and contact angle are plotted for comparison. This relationship is known from simple geometrical considerations to be

$$R = \sin\delta \left( \frac{3V}{\pi\, 2 - 3\cos\delta + \cos^3\delta} \right)^{1/3}, \qquad (2)$$

with $R$ the drop radius. (Although a spherical drop with a circularly shaped contact line is an approximation, it serves our purpose here, as a hexagonal or triangular drop has 6-fold or 3-fold rotational symmetry, so that the deviation is not too large.)

As can be clearly seen in figure 5(a), a single printed drop is initially overstretched considerably so that its effective contact angle is $\sim 10°$, whereas the intrinsic contact angle of the surface is $63°$. The initial overspreading of the drop is due to the excess kinetic energy as it impacts the surface, and then pinning by the posts stops the interface retracting. For larger drops, formed from an increasing number of ink-jetted drops, the data in figure 5(a) shows that the effective contact angle rises to $120^0$. Indeed close analysis of the microscopy images confirms that the spherical cap of the large drops overhangs beyond the contact line on the surface, indicating that the contact angle exceeds $90°$. This occurs because the net available energy at the contact line per jetted drop decreases over time, as increasingly this energy will be absorbed by the fluid reservoir of the main drop and dissipated viscously. There is not enough energy to cause further spreading, and the contact line remains pinned. Hence the effective contact angle, rather than the footprint of the drop increases.

From the geometrical parameters of the drop we can estimate that for a single 200pl drop, the average height is $h \approx 5\mu m$ corresponding to about one-third of the height of the posts. For the larger drops the liquid-vapour interface forms an obvious spherical cap over the posts. Figure 5(b) indicates that, despite this difference in geometry and in final effective contact angle, the contact line shapes are very similar: the average of the parameter $\lambda$ is $\langle\lambda\rangle = 0.55$, with a standard deviation of only $\sigma_\lambda = 0.06$. Thus pinning, and the preference of the contact line for depinning via the connected contact line mechanism, are valid not only for liquid-vapour interfaces which wet up to the post height, but also for interfaces that wet only part of the pillars or extend above the pillars. Moreover the final drop imprints remain qualitatively the same in the experiments, where large amounts of kinetic energy are available, and in the simulations which are quasi-static.

## 4. Receding contact line motion
### 4.1. Experimental results

In the second part of this paper we compare experiments and simulations aimed at understanding the behaviour of a drop as it evaporates from a surface patterned with polygonal posts. We start with the anisotropic, pinned drops, created using ink-jet printing, and described in section 3.

A typical example of the behaviour of an evaporating drop as a function of time is shown in figure 6(a). During evaporation of the water drops, the spherical cap of liquid above the posts starts to evaporate while the outer contact line is observed by microscopy to be pinned. The contact line typically remains pinned until the tops of the outermost posts are de-wetted. This de-wetting can be clearly observed during the experiment due to the sudden contrast change of the posts. It is important to note that the retraction process does not necessarily start after all tops are de-wetted but, typically, while some of the tops are still wet. The drop then retracts post by post (or several posts at a time) with a jumping motion of the contact line, until only a very small group of posts remains wet. At this point, the drop evaporates quickly and completely. During de-wetting, the contact line changes in shape from the initial irregular hexagon to a round shape. Normally the drop shape does not change into a regular hexagon before becoming rounded. However, varying irregular shapes can also form during the evaporation, as symmetry is often broken by one part of the contact line starting to move inwards while another remains pinned. These irregular shapes suggest that kinetics and surface defects play dominant roles for small liquid volumes combined with high evaporation rates.

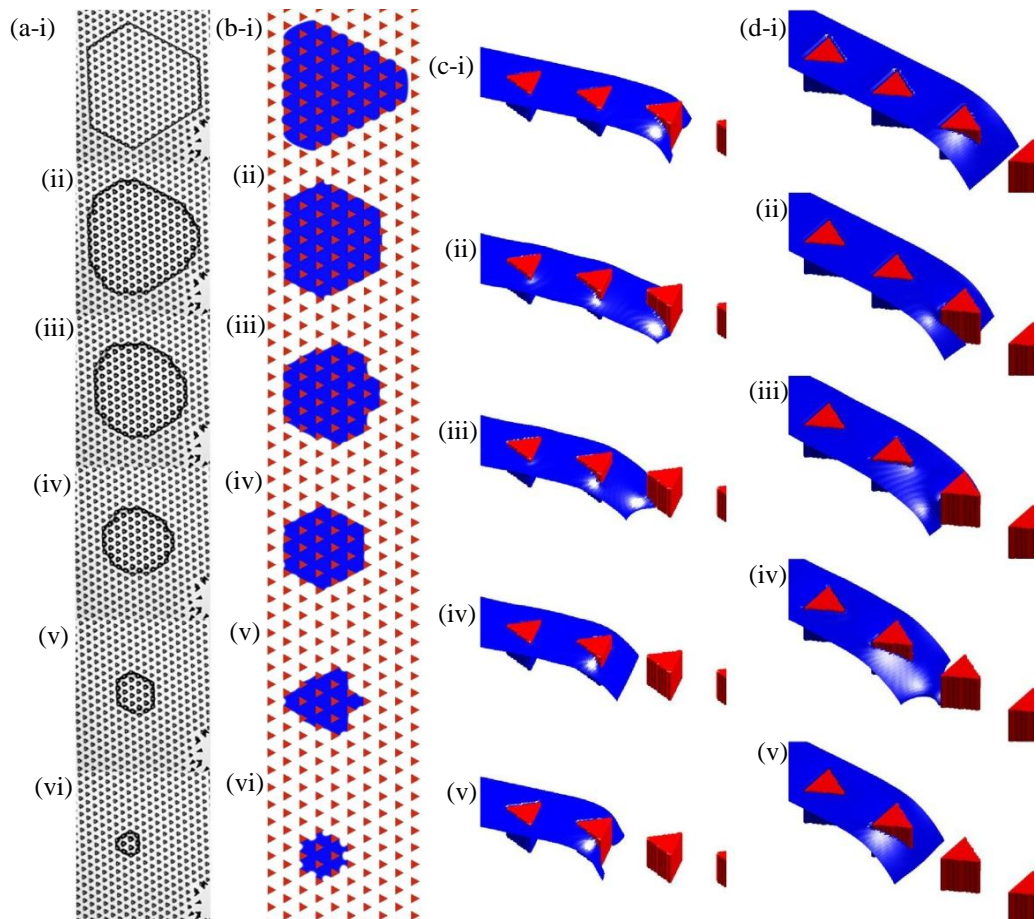

**Fig. 6** (a) Experimental and (b-d) simulation results showing receding contact lines on surfaces patterned with triangular posts on a hexagonal lattice (with $\delta = 55°$, h=30, b=20, h=40). The roman numerals indicate increasing time. (a) Time sequence microscopy images of the evaporating drop. The black outlines indicate ,positions of the contact line at various times during evaporation. (b) 3D lattice Boltzmann simulation, employing quasi-static evaporation, where (i) is the starting position (initial pinned contact line), and (vi) is the final drop position before complete evaporation. The contact line retracts in a very similar way to the experiments in (a). (c) & (d) Simulations of (c) disconnected contact line (DCL) de-pinning and (d) a mixed connected contact line (CCL) (i-ii) and then DCL de-pinning (iii-v) for a receding contact line. At (v) the contact line has moved exactly one lattice spacing from (i). The simulations are periodic in a direction perpendicular to the motion of the interface.

### 4.2. Simulation results

To investigate the effects of patterning anisotropy on evaporation by means of lattice Boltzmann simulations we take an imbibed film as the starting point in our simulation. In particular, we begin with the configuration in figure 3(c), obtained by quasi-static spreading on a hexagonal lattice of triangular posts with intrinsic contact angle $\delta = 55°$. Rather than using a virtual drain to drive the drying, we gradually reduce the volume of the film, by subtracting a small amount of mass from the liquid phase at each time step, such that quasi-static evaporation is simulated. This corre-sponds more closely to the experimental situation.

The evolution of the simulated drop in figure 6(b) is now compared to the experimental results. The main trend is that the film retracts preferentially towards the points of the triangles compared to the faces. This effect is sufficiently strong to undo, and indeed reverse, the anisotropy gained in the spreading phase. One of the intermediate contact line contours, 6(b-iv) is a nearly regular hexagon, indicating that the drop reverts back to this original shape during evaporation, before proceeding to smaller shapes where the drop footprint is a triangle pointing left, rather than right. This is similar to the behaviour seen in the experiments, although the faceting is less pronounced in the latter case. (The simulation suffers from a small anisotropy resulting from the cubic discretisation used in the lattice Boltzmann algorithm, which breaks the natural three-fold symmetry of the physical system, so that retraction in the $x$ direction is slightly slower than that in the two diagonal directions, causing the broken symmetry seen in 6(b-iii) and (b-v).)

We now look at the possible de-pinning mechanisms in detail. These are summarised in figures 6(c) and 6(d). As for the advancing contact line case, the simulations are periodic in a direction perpendicular to the motion of the interface.

Consider first the case where the retreating interface reaches the face of the triangular posts first, shown in figure 6(c). In the step (i) to (ii) in the figure, the contact line remains pinned to the bottom corners of the post while the interface de-wets the sides of the post vertically. Once the height of the interface around the boundary post has dropped significantly, the contact line de-pins from the corners of the posts, but (iii) demonstrates that there is still a barrier to retraction, with the interface being significantly distorted around the post. This barrier can be understood in terms of the free energy cost of increasing interfacial area in the widening gap. When the contact line finally reaches the apex of the triangle, it snaps back into a straight, connected configuration, as shown in (iv). The contact line then quickly retracts, readily de-wetting the front face of the next post, until it is in a disconnected state as shown in (v), equivalent to its starting configuration in (i). The interface has now moved through one lattice spacing and the cycle repeats. This mechanism can be considered as disconnected contact line (DCL) de-pinning, as the front face of the post has to be de-wet first forming a disconnected contact line. This requires considerable distortions of the liquid-vapour interface, and it is analogous to DCL de-pinning for advancing contact lines.[30,31]

The snapshots in figure 6(d) show the depinning mechanism when the retreating interface reaches an apex of a triangular post first. From the connected configuration in (i), the contact line retreats until it is punctuated by the posts in (ii). Then the liquid de-wets the two front faces of the posts, so that the contact line is level with the back of the posts, as shown in (iii). As (iv) demonstrates, de-wetting the back face of the post is the main obstacle to de-pinning, with the interface being greatly distorted. Once the interface has de-pinned from this post, the contact line springs back into position, (v), equivalent to position (i) except for a lattice translation, completing the cycle. This is the first de-pinning mechanism that we have discussed where pinning occurs both when the contact line is connected (d(i)-d(ii)) and when it is disconnected(d(iii)-d(iv)), so it is best seen as a mixed CCL/DCL route. This is in contrast to the advancing case where the contact line is only pinned when it is connected.

To better understand the evaporation process, we next use similar simulations to investigate both mechanisms of de-pinning simultaneously during retraction of the film, as shown in figure 7. As before we consider a geometry where the simulations are periodic in a direction perpendicular to the motion of the interface. We initialise the system with a strip of liquid filling the substrate to the tops of about 7 posts and again model evaporation by slowly subtracting mass.

Figure 7(a) shows the simulation results for the time evolution of the film for a contact angle $\delta = 30°$, close to the the value of the receding contact angle for the experimental surfaces. For this low value of $\delta$, de-pinning from the outer posts is inhibited in *both* directions. Instead, the film decreases in thickness as it evaporates, while remaining at a constant length. This is a consequence of the increased energy cost of de-wetting a more hydrophilic substrate. The film only begins to retract when it is very thin: less than half of $h$. As before it retracts only in the direction of the CCL/DCL mixed de-pinning, which indicates that the preferred de-pinning direction persists for films thinner than the post height.

The situation for $\delta = 55°$ is shown in figure 7(b). For this contact angle the liquid thickness remains constant at $h$, and the preference is again for CCL de-pinning. Increasing $\delta$ up to $70°$ gave no qualitative change. In general, the contact angle sensitivity for retraction appears to be less than that for advancing contact lines. We were unable to identify a regime where DCL depinning occurred in preference to mixed CCL/DCL retraction. This is different to the advancing contact line case where Blow and Yeomans[31] showed that either CCL and DCL de-pinning may be preferred depending on the contact angle and the ratio of post height and separation.

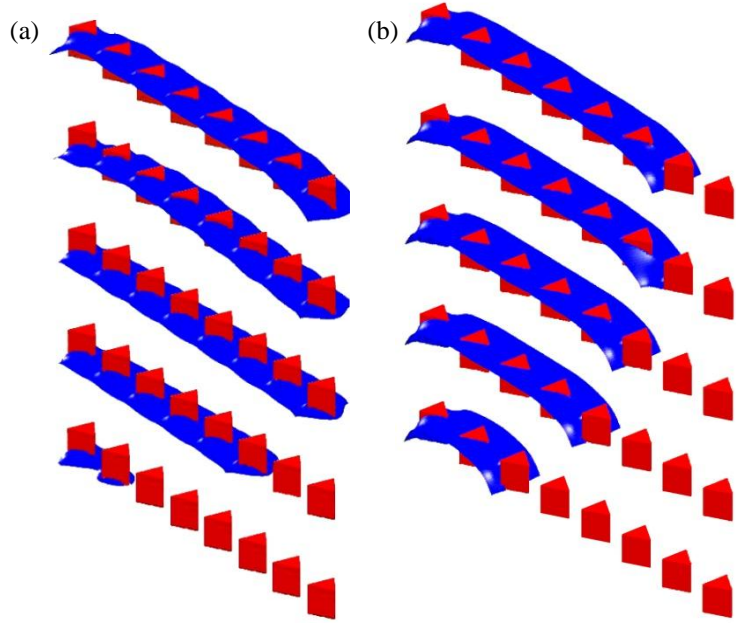

**Fig. 7** Receding behaviour of a drying lattice Boltzmann drop on a surface of posts (*with b = 20, d = 40 and h = 20*) without mirror symmetry, for increasing time from top to bottom. The simulations are periodic in a direction perpendicular to the motion of the interface. (a) For $\delta = 30°$, the drop remains pinned until the tops of the posts are de-wetted. Then, the interface de-pins from the face of the triangle. (b) For $\delta = 55°$, de-pinning from the face occurs before de-pinning from the top of the interface and the drop retains constant height away from the contact line.

### 4.3. Comparison and reproducibility

There are two main differences between the experimental and computational setups. Firstly, in the simulations we modelled an imbibed film lying entirely amidst the posts. This corresponds in the experiments to the case of a single or few inkjet-printed drops, but not to the case of multiple drops where the interface forms a dome-shaped cap above the pillars. However, during evaporation it is observed with microscopy that the spherical cap shrinks while the contact line remains pinned, until the tops of the posts are dewetted, at least at the periphery. Therefore, the starting position of the simulations should provide a good model.

Secondly, and more importantly, there are many more posts under the drop in the experiments than in the simulations. This makes a quantitative comparison infeasible. In particular, in Figure 6(a), the receding contact line appears increasingly circular as evaporation progresses. This is not witnessed in the simulations, where the contact line remains more obviously facetted. This is probably because the small number of posts under the simulation drop gives limited scope to form a circular shape. Nevertheless, the two methods clearly agree on the qualitative nature of the anisotropy; that retraction is energetically favoured for contact lines along which the points of the triangular posts point outwards. Since an equivalent CCL de-pinning mechanism also plays the dominant role during imbibition (advancing contact line), the drops preferentially advance and retract along the same lattice directions, leading to the initial drop shape being approximately recovered as the contact line retracts.

On ideal surfaces, a drop will retract to its centroid, $M$ in figure 4. For the experimental drops $z$, the actual position of the last post to de-wet, was measured in the three lattice directions. $M$ can be calculated from the experimental data, assuming the geometry of figure 4(b), via

$$M = \frac{K}{2}\tan(30°) = \frac{x+y}{3} = \frac{2y}{3}\frac{1}{2-\lambda}. \qquad (3)$$

Averaging over data for drops of different initial volumes gave mean values $\langle z/y \rangle$ and $\langle M/y \rangle$ of 0.45 and 0.46 respectively, and there was no obvious difference in the means for each of the three lattice directions. This shows that the system on average retains its three-fold rotational symmetry and indeed retracts towards the centroid on average. However, the scatter in the experimental data was considerably larger than in the advancing case, with standard deviation $\sigma_{z/y} = 0.15$. The simulation results were $\langle z/y \rangle_{sim} = 0.39$, $\sigma_{z/y,sim} = 0.10$, which is also well within the predicted range. The spread in the simulations, along with the fact that the drop does not retract exactly to its centroid, were due in part to lattice discretisation errors, which were exacerbated by the small number of posts initially covered.

We found that the drop shapes were much more reproducible under spreading than under retraction, as evidenced by the higher scatter in $\sigma_{z/y}$ than in $\sigma_\lambda$. Retraction takes place through an obvious unzipping mechanism: the dewetting of one pillar is often followed by a neighbouring row becoming de-wet. Any local variations in post size or spacing may nucleate such a mechanism which can have a large impact on the overall drop shape. We also note that dynamical effects increase with time for evaporation, since the evaporation rate proportional to volume grows larger, while for spreading the kinetic energy dissipates with time.

## 5. Conclusions

In this paper we have compared experiments and lattice Boltzmann simulations to gain an understanding of the shapes of ink-jet printed drops on surfaces patterned with a regular lattice of polygonal posts. We considered diamond and, in more detail, triangular posts with post widths typically ~7μm and lattice spacings between 15 and 40μm. Complex drop shapes with 3, 4 and 6-fold symmetry were observed. We drew on evidence from the experiments and simulations to argue that the final shape of a spreading drop depends on both the symmetry and spacing of the post lattice, and on the shape and dimensions of the posts. This is because the shape is controlled by the interface pinning on the edge of the posts. Anisotropic drop shapes result when the range of contact angles over which the interface pins is different in different lattice directions: typically this occurs when the ratio of post width to spacing lies between 0.25 and 0.75.

The lattice Boltzmann simulations allowed us to describe the different pinning mechanisms in detail for triangular posts, and hence to explain why the interface spreads most easily in the directions when it is moving towards the flat faces of the triangular posts.

Experiments were performed both for single drops and for larger drops which were produced by the sequential jetting of up to 255 drops at the same position. The final, anisotropic, footprint of the drop was very similar for the different drop volumes, and for the quasi-static simulations, demonstrating the robustness of the pinning mechanisms. Rather than causing the drop the spread further across the surface, increasing the amount of fluid increases the final effective contact angle and, for larger drops, a spherical cap was formed above the posts.

We then performed experiments and simulations to study evaporation for the same systems. The portion of the drop above the posts starts to evaporate first, and then the fluid lying within the posts starts to retract. As the drop retracts it tends to regain a circular shape centred at the initial point of impact. To explain this we used the simulations to identify the mechanisms for interface depinning, showing that for both the advancing and the receding case the easy direction is the same. Hence, as a drop retracts, it tends to regain its original shape.

It would be interesting to perform further experiments focussing on the link between the spreading, the velocity and the print frequency of the incoming ink drops. Another interesting avenue for further study is the interplay of the evaporation from above and between the posts. Although our study addresses fundamental issues of drops spreading on complex surfaces, it suggests that polygonal posts may be exploited to control liquid motion. They may, for example, lead to new design options in micro-fluidic devices. Indeed there is in principle no need for a regular lattice, and non-periodic placement of posts could lead to new drop shapes. By also locally varying the post shapes or the post orientation relative to the lattice, yet more design freedom could be created.

The agreement between lattice Boltzmann simulations and experiments using inkjet printing is particularly encouraging. On one hand, it shows the strength of using the lattice Boltzmann method to guide experiments on wetting phenomena. On the other hand, inkjet printing was shown to be a viable technique for continued research in this direction, leading to reproducible results for drops as small as 10 picoliter.

## Acknowledgements

MLB acknowledges the support of the Portuguese Foundation for Science and Technology (FCT), through the grants SFRH/BPD/73028/2010 and PEst-OE/FIS/UI0618/2011, and JMY the support of the ERC through the Advanced Grant MiCE.

## Supporting material: simulation details

We model the system as a diffuse-interface, two-phase fluid in contact with a solid substrate. The thermodynamic state of the fluid is described by an order parameter $\rho(r)$, corresponding to the density of the fluid at each point $r$. The equilibrium properties are modelled by a Landau free energy functional over the spatial domain of the fluid $D$, and its boundary with solid surfaces $\partial D$, as [1]

$$\Psi = \iiint_D (p_c\{v^4 - 2\beta\tau_W(1-v^2)-1\} - \mu_b\rho + \frac{1}{2}\kappa|\nabla\rho|^2)dV - \iint_{\partial D} \mu_s\rho dS. \tag{1}$$

The first term in the integrand of (4) is the bulk free energy density, where $v = (\rho - \rho_c)/\rho_c$ and $\rho_c$, $p_c$, and $\beta\tau_W$ are constants. It allows two equilibrium bulk phases, liquid and gas, with $v = \pm\sqrt{\beta\tau_W}$. The second term is a Lagrange multiplier constraining the total mass of the fluid. The third term is a free energy cost associated with density gradients. This allows for a finite width, or diffuse, interface to arise between the bulk phases, with surface tension $\gamma = \frac{4}{3}\rho_c\sqrt{2\kappa p_c(\beta\tau_W)^3}$ and width $\chi = \frac{1}{2}\rho_c\sqrt{\kappa(\beta\tau_W p_c)^{-1}}$. The boundary integral takes the form proposed by Cahn.[2] Minimizing the free energy leads to a Neumann condition on the density

$$\partial_\perp\rho = -\frac{\mu_s}{\kappa}. \tag{2}$$

The wetting potential $\mu_s$ is related to the Young angle $\delta$ of the substrate by [1]

$$\mu_s = 2\beta\tau_W\sqrt{2p_c\kappa}\;\text{sign}(\frac{\pi}{2}-\delta)\sqrt{\cos\frac{\alpha}{3}(1-\cos\frac{\alpha}{3})} \quad \text{with} \quad \alpha = \arccos(\sin^2\delta). \tag{3}$$

The hydrodynamics of the fluid is described by the continuity and the Navier-Stokes equations

$$\partial_t\rho + \partial_\alpha(\rho u_\alpha) = 0 \tag{4}$$

$$\partial_t(\rho u_\alpha) + \partial_\beta(\rho u_\alpha u_\beta) = -\partial_\beta P_{\alpha\beta} + \partial_\beta(\rho\eta(\partial_\beta u_\alpha + \partial_\alpha u_\beta) + \rho\lambda\delta_{\alpha\beta}\partial_\gamma u_\gamma), \tag{5}$$

where $u$ is the local velocity, $P$ is the pressure tensor derived from the free energy functional (1) as

$$P_{\alpha\beta} = \left(p_c(v+1)^2(3v^2 - 2v + 1 - \beta\tau_W) - \frac{1}{2}\kappa\partial_\gamma\rho\partial_\gamma\rho - \kappa\rho\partial_{\gamma\gamma}\rho\right)\delta_{\alpha\beta} + \kappa\partial_\alpha\rho\partial_\beta\rho, \tag{6}$$

with $\eta = \frac{\Delta t}{3}(\tau - \frac{1}{2})$ and $\lambda = \eta(1 - 12p_c\rho_c^{-2}\rho(3v^2 - \beta\tau_W))$ are the shear and bulk kinematic viscosities respectively. A free energy lattice Boltzmann algorithm is used to numerically solve equations 7 and 8.[3-5] At the substrate we impose the boundary condition (5),[1,6] and a condition of no-slip.[7-9] We choose $\kappa = 0.01$, $p_c = 0.125$, $\rho_c = 3.5$, $\tau_W = 0.3$ and $\beta = 1.0$, giving an interfacial thickness $\chi = 0.9$, surface tension $\gamma = 0.029$ and a density ratio of 3.42. The viscosity ratio is $\eta_L/\eta_G = 7.5$.

We concentrate our attention on substrates patterned with hexagonal arrays of posts with equilateral triangles as cross-sections. The posts have height $h = 20\,l.u.$ (lattice units or cells), side length $b = 20\,l.u.$, and centre-to-centre separation $d = 40\,l.u.$. The posts and the base substrate have the same Young angle $\delta$.